\documentstyle[aps,prb,preprint]{revtex}
\tighten
\begin{document}
\title{Quantum critical point and scaling in a \\ layered array of ultrasmall
Josephson junctions.}
\date{\today}
\author{T. K. Kope\'{c}$^{1,2}$ and J. V. Jos\'{e}$^{1}$}
\address{$^{1}$Physics Department and Center for Interdisciplinary
Research on Complex Systems, Northeastern University,
Boston, MA 02115 USA\\
$^{2}$Institute for Low Temperature and Structure Research,
Polish Academy of Sciences,\\
 POB 1410, 50-950 Wroclaw 2, Poland}
\maketitle
\begin{abstract}
We have studied a quantum Hamiltonian that models an array of ultrasmall
Josephson junctions with short range Josephson couplings, $E_J$, and
charging energies, $E_C$,  due to the small capacitance of the junctions.
We  derive a new  effective quantum spherical model for the
array Hamiltonian. As an application 
we start by approximating the capacitance matrix by its self-capacitive 
limit and in the presence of an external uniform  background of charges, $q_x$.
In this limit we obtain the zero--temperature
superconductor--insulator phase  diagram,  $E_J^{\rm crit}(E_C,q_x)$,
that improves upon previous theoretical results  that used a mean field theory 
approximation. Next we  obtain a closed--form expression for the 
conductivity of a square array, and derive  a universal 
scaling relation  valid about the zero--temperature quantum 
critical point. In the latter regime the energy scale is determined 
by temperature and we establish universal
scaling forms for the frequency dependence of the conductivity.

\end{abstract} 

\newpage

\section{Introduction}
Quantum phase transitions have attracted a significant amount of
interest in recent years. There are different physical systems where quantum
phase transitions can be studied, like in magnetic systems
\cite{mag}, the quantum Hall effect \cite{qhe}, superconducting 
films \cite{films} and Josephson junction arrays (JJA) 
\cite{reviews}. There have been significant advances in lithographic 
techniques  that allow the fabrication of  arrays of ultra-small 
superconducting  islands, with charging energy, $E_C$, that can dominate the
Josephson coupling energy, $E_J$,  making  quantum fluctuation effects
of paramount importance \cite{simanek}.  Because the junction parameters 
can be controlled  accurately JJA offer a unique system where one can  
test the nature of quantum phase transitions and critical phenomena,
in particular the superconductor--insulator (SI) phase transition 
induced by quantum fluctuations. 
Therefore JJA can be a prototype system that can display well controlled 
quantum SI phase transitions.

In the JJA problem, the two relevant low temperature energy scales 
are $E_J$, that
permits tunneling of Cooper pairs between islands, and  $E_C$ that 
tends to localize the charge carriers in the islands.
When $E_J$ is much larger then $E_C$ the phases in the islands 
are well defined. In this regime only the phases determine the 
properties of the JJA. In the opposite limit, i.e.
$E_C>>E_J$ the superconducting phase is perturbed by strong zero point
quantum fluctuations due to the Coulomb blockade that localizes the charge
carriers to the islands. Several theoretical ~\cite{simani2,eckern,rojas} 
and experimental studies~\cite{zant}  have considered 
the competition between the $E_J$ dominated phase  
and the $E_c$ dominated charging energy regions
in periodic JJA. It has been established that for 
sufficiently large charging energy the quantum phase fluctuations lead 
a complete suppression of long--range phase coherence 
even at zero temperature. This type of 
{\it quantum phase transition}  has attracted 
significant interest in recent
years (for a review see Ref.\onlinecite{sondhi} and also \onlinecite{subir}).
 Generally, these transitions take place at zero temperature
where crossing the phase boundary is accompanied by a change
in the ground state of the system. This transition is induced 
at zero temperature by changing some external 
parameter in the Hamiltonian of the system, for example the 
charging energy in the  Josephson-junction arrays (given by the
quantum parameter $\alpha\equiv E_C/E_J$). 
Let the critical temperature be $T_c$ and denote the
distance from criticality as $\delta t=1-T/T_c$.
The critical behavior is  asymptotically close to 
the critical point which is entirely determined by classical 
physics.  This is because the characteristic frequency associated 
with the critical fluctuations, $\omega_c\sim 1/\xi_\tau$, vanishes at
criticality, and the characteristic correlation time for
relaxation towards equilibrium $\xi_\tau\sim \xi^z$ diverges at 
the critical point (here $\xi\sim |t|^{-\nu}$
is the diverging characteristic correlation length in the system.)
A quantum system behaves classically if the temperature is larger than all
frequencies of interest, i.e. as long as $\hbar\omega_c<<k_BT$,
which is always the case when the system is sufficiently close 
to the critical point with a non zero $T_c$. 
In the critical region various physical quantities can have  scaling forms
(i.e. given in terms of homogeneous functions) 
as a consequence of the diverging $\xi$ and $\xi_\tau$. 
Since the experimentally accessible
systems are necessarily at  non-zero temperature,
one needs to understand how the quantum behavior  is modified at
finite temperature. The important observation is that at finite temperature 
the partition function has a finite temporal dimension, since
the effective classical system extends to an  extra dimension
of size $\beta\hbar$.

At $T=0$,  the 2D JJA problem goes into a
 $3$--dimensional classical XY model with
the critical behavior of the 3D XY model.
Moreover, there should be a clear signature in the nature of the
correlation functions (like for example in linear response)
when crossing over from the 2D XY model at high temperatures to the 3D XY
model as $T\to 0$. Specifically, the response of the
quantum critical point to a small temperature perturbation
will be determined by the location of the quantum--to--classical
crossover. An especially interesting {\it quantum critical regime}
appears when the relaxational frequency satisfies $\hbar\omega\approx k_BT$.
In this regime the only relevant scale is given by the temperature 
and the dynamical scaling functions will only depend on the ratio  
$\hbar\omega/k_BT$. As the ratio $\hbar\omega/k_BT$ is varied in 
an experiment we expect to see a crossover between temperature-- 
and frequency--dominated scaling regimes. For the finite temperature
experiments one can scale the data by  using the knowledge of the 
frequency dependent  scaling  functions to test the existence or
probing of a quantum critical  point.
The experimental identification of a quantum  phase transition in JJA
will rely upon finding the scaling behavior with the relevant parameters as
temperature, frequency and  the wavelength dependence of various observables.
The specific signatures are given by universal values of certain dimensionless
amplitudes as,  for example, the resistivities at the critical 
point in 2D JJA systems. Motivated by these issues, it is the 
goal of the present paper to  study the superconductor--insulator 
transition in 2D JJA in the quantum critical regime, by focusing 
on the  scaling properties of the directly measurable 
electromagnetic conductivities in JJA systems.

The outline of the rest of the paper is the following:
In Section II we define the model Hamiltonian, followed by a novel
path integral formulation of the problem. In section III we present
as a test of our quantum spherical model approach for the self-capacitance
model.  There we show that the spherical technique gives a better answer 
than the one obtained from mean field theory. In Section IV we present
our calculation of the scaling properties of the conductivity in the quantum
critical regime.  Finally, in Section V we briefly discuss our results.

\section{The model}

\subsection{Quantum phase Hamiltonian}
A Josephson junction array (JJA) can be modeled by a periodic lattice
of superconducting islands separated by insulating
barriers. Each island becomes independently superconducting about the bulk
transition temperature $T_{c0}$ and it is characterized by an
order parameter $\psi({\bf r}_i)=|\psi_0({\bf r}_i)|e^{i\phi({\bf r}_i)}$, 
where ${\bf r}_i$ is a two--dimensional vector denoting the position 
of each island. The magnitude of the order parameter, $|\psi_0({\bf r}_i)|$, 
is non--fluctuating when the temperature is lowered further 
and the onset of long range phase coherence due to the tunneling 
of Cooper pairs between the  islands is responsible for the zero 
resistance drop in the arrays.  The phase coherence onset temperature in JJAs
can be significantly reduced by making the junctions ultrasmall.
When the junction's capacitance is small the charging  energy
(i.e. the energy necessary to transfer Cooper pair charges between the islands)
increases to the point where no pairs can tunnel anymore, completely quenching
the Josephson current. The competition  between the Josephson tunneling
and the charging (Coulomb) energy, without dissipation, 
can be modeled by the Hamiltonian
\begin{eqnarray}
H&&=H_C+H_J,
\nonumber\\
H_C&&=-\frac{1}{2}\sum_{{\bf r}}
[{\bf C}^{-1}]_{{\bf r}{\bf r}'}{\hat Q}_{\bf r}{\hat Q}_{{\bf r}'},
\nonumber\\
H_J&&=\sum_{\langle {\bf r}_1,{\bf r}_2\rangle}
J(|{\bf r}_1-{\bf r }_2|) \left[1-\cos(\phi_{{\bf r}_1}-
\phi_{{\bf r}_2})\right].
\label{hamil}
\end{eqnarray}
\narrowtext
Here, $\hat{Q}_{\bf r}=({2e}/{i}){\partial}/{\partial\phi_{\bf r}}$ is the
charge operator while ${\hat \phi}_{\bf r}$  represents 
the superconducting phase operator of the grain at the site $\bf r$;
$J(|{\bf r}_1-{\bf r }_2|)$ is the site--dependent
Josephson coupling  and  $[{\bf C}^{-1}]_{{\bf r}{\bf r}'}$
is the inverse capacitance matrix.

\subsection{Euclidean action and path integral representation}

It is useful to derive a field-theoretic representation of 
the partition function for Eq (1) to study the quantum 
nature of the superconductor-insulating (SI) phase transition.
A convenient procedure is to introduce a path--integral
representation in a basis diagonal in $\phi_j$.
In this representation the partition function is expressed as
\begin{equation}
Z=\int\left[\prod_{\bf r}{\cal D}\phi_{\bf r}\right]e^{-{\cal S}[\phi]}.
\label{partfun}
\end{equation}
Here the functional integral is evaluated over the phases restricted
to the compact interval $[0,2\pi]$ and
with an effective action ($\hbar=1$)
\begin{eqnarray}
{\cal S}[\phi]&=&{\cal S}_C[\phi]
+{\cal S}_J[\phi],\nonumber\\
{\cal S}_C[\phi]&=&\frac{1}{8e^2} \int_0^\beta d\tau
\sum_{\bf r,r'}
[{\bf C}^{-1}]_{{\bf r}{\bf r}'}
\left(\frac{\partial\phi_{\bf r} }{\partial\tau}
\right)
 \left(\frac{\partial\phi_{\bf r'} }{\partial\tau}
\right),
\nonumber\\
{\cal S}_J[\phi]&=&\int_0^\beta d\tau
\sum_{\langle {\bf r}_1,{\bf r}_2\rangle}
J(|{\bf r}_1-{\bf r }_2|) \left\{1-\cos[\phi_{{\bf r}_1}(\tau)-
\phi_{{\bf r}_2}(\tau)]\right\}.
\label{action}
\end{eqnarray}

Because the values of the phases which differ by $2\pi$ are equivalent,
 the path integral can be written in terms of
non--compact phase variables $\theta_{\bf r}(\tau)$, 
defined on the unrestricted
interval $[-\infty,+\infty]$, and by a set
of winding numbers $\{n_{\bf r}\}$, which are integers
running from $-\infty$ to $+\infty$.
Consequently, the path--integral representation 
in Eq.(\ref{partfun}) also includes
a summation over winding numbers\cite{Bruder}:
 $\phi_{\bf r}(\tau)=\theta_{\bf r}(0)
+2\pi i n_{\bf r}\tau/\beta +\theta_{\bf r}(\tau)$ which
reflects the discreteness of the charge, so that the integration measure reads
\begin{equation}
\int{\cal D}\phi_{\bf r}\equiv
\sum_{\{n_{\bf r}\}}\prod_{\bf r}
\int_0^{2\pi} d\theta(0)\int_{\theta(0)}^{\theta(0)+2\pi n_{\bf r}}
{\cal D}\theta_{\bf r}(\tau).
\end{equation}

This concludes the formulation of the model. In the following
section we will transform this representation to one where 
the order parameter field is expressed in a novel 
quantum spherical model self--consistent scheme.

\section{Quantum spherical model approach}

To study the JJA model it appears at first natural
to use a description in terms of an effective
Ginsburg--Landau functional derived from the microscopic model
of Eq. (\ref{action}). Several studies of JJA have followed this route,
also known as the coarse grained approach first developed by Doniach
\cite{doniach}. The key point of this method is to introduce
a complex field order parameter ${\mbox{\boldmath$\psi$}}_{\bf r}$
(or equivalently a two component real field) whose expectation value
is proportional to the ${\bf S}_{\bf r}(\phi)$
vector defined by
\begin{equation}
{\bf S}_{\bf r}(\phi)
=[S^x_{\bf r}(\phi),S^y_{\bf r}(\phi)]
\equiv [\cos(\phi_{\bf r}),\sin(\phi_{\bf r})].
\label{spin}
\end{equation}
The non-zero thermal average $\langle S^x(\phi)  \rangle=\langle\cos\phi_i   \rangle$
describes the ``phase-locking" or long--range phase ordering in the model.
The system is governed by the Ginzburg-Landau functional as long as the order parameter is small
and the decoupling of the Josephson energies is valid.
This is a serious shortcoming of the coarse grained approach since
the method is restricted to the region of parameters in the
vicinity of the critical point and does not offer
a self--consistent description of the full problem.

There are no nontrivial exact solutions of the model, 
except for some molecular field--type approaches. It is therefore
reasonable to develop new approximate mappings of the model
which may admit exact solutions, both in the ordered in disordered phases.
>From the definition of the pseudo--spin
variables $S_{\bf r}$ the following rigid constraint holds for each one, 
\begin{equation}
|{\bf S}_{\bf r}(\phi)|^2=[S^x_{\bf r}(\phi)]^2+[S^y_{\bf r}(\phi)]^2=
\cos^2(\phi_{\bf r})+\sin^2(\phi_{\bf r})\equiv 1.
\label{spin2}
\end{equation}
The relation in Eq. (\ref{spin2}) also implies that a weaker condition
also applies, namely:
\begin{equation}
\sum_{\bf r}|{\bf S}_{\bf r}(\phi)|^2=N.
\label{constr}
\end{equation}
The main idea of our approach is to attempt to generate an effective partition function
from the original one with cosine interaction,
which  incorporates the constrained nature of the original variables.
This leads us to the
formulation of the problem in terms of the spherical model~\cite{spher,spher2,spher3}
by introducing the appropriate constrained order parameter field.
The name of the model comes from the observation that in Eq.(\ref{constr})
the allowed states
of the spherical model are all points on the surface of a hypersphere
of radius $N$. Although the spherical model was originally introduced as
an approximation to the classical Ising model, the ``sphericalization"
technique may be readily applied to a variety of other problems
(for a quantum generalization of the spherical model,
see Ref.\onlinecite{qspher});
the one key necessary ingredient is the existence of an inherent constraint on
the Hamiltonian variables as in Eq.(\ref{spin}). 
The model defined by Eq.(\ref{constr}) is in fact a hybrid of the spherical
and the two component ($M=2$) vector models. The global constraint of Eq.
(\ref{constr}) may be introduced on the continuous order parameter field
by using the
functional analogue of the Dirac--delta function 
\begin{eqnarray}
1&\equiv&
   \int\left[\prod_{\bf r} {\cal D}
{\mbox{\boldmath$\psi$}}_{\bf r}{\cal D}
{\mbox{\boldmath$\psi$}}^\star_{\bf r}\right]
\delta\left(\sum_{\bf r}|{\mbox{\boldmath$\psi$}}_{\bf r}(\tau)|^2-N\right)
\nonumber\\
&&\prod_{\bf r} 
\delta\left[{\mbox Re}\,{\mbox{\boldmath$\psi$}}_{\bf r}(\tau)-
{ S}_{\bf r}^x(\phi(\tau)
\right]
\delta\left[\mbox{Im}\,{\mbox{\boldmath$\psi$}}_{\bf r}(\tau)-
{ S}_{\bf r}^y(\phi(\tau))
\right],
\end{eqnarray}
where the ${\mbox{\boldmath$\psi$}}_{\bf r}(\tau)$ are the
complex $c$--number  fields, which satisfy the quantum
periodic boundary condition
${\mbox{\boldmath$\psi$}}_{\bf r}(\beta)=
 {\mbox{\boldmath$\psi$}}_{\bf r}(0)$, and
taken as continuous variables, i.e.,
$-\infty  <   {\mbox{\boldmath$\psi$}}_{\bf r}(\tau)  <+\infty$,
but constrained (on the average) to have unit length.
The partition function of Eq. (\ref{partfun}) then reads
\begin{eqnarray}
Z&=&\int\left[\prod_{\bf r} {\cal D}{\mbox{\boldmath$\psi$}}_{\bf r}
{\cal D}{\mbox{\boldmath$\psi$}}^\star_{\bf r}\right]
\delta\left(\sum_{\bf r} |{\mbox{\boldmath$\psi$}}_{\bf r}(\tau)|^2
-N\right)
e^{-{\cal S}_J[{\mbox{\boldmath$\psi$}}]}\times
\nonumber\\
&\times&\int\left[\prod_{\bf r}{\cal D}\phi_{\bf r}\right]
e^{-{\cal S}_C[\phi]}
\prod_{\bf r} 
\delta\left[{\mbox Re}\,{\mbox{\boldmath$\psi$}}_{\bf r}(\tau)-
{ S}_{\bf r}^x(\phi(\tau)
\right]\nonumber\\
&\times&
\delta\left[{\mbox Im }\,{\mbox{\boldmath$\psi$}}_{\bf r}(\tau)-
{ S}_{\bf r}^y(\phi(\tau))
\right].
\end{eqnarray}
One may wonder whether the introduction of the spherical
condition on  the order parameter fields
in Eq. (\ref{constr}) is essential
since the global constraint is automatically fulfilled
by virtue of the exact relation (\ref{spin2}).
 Indeed, in a rigorous  treatment
of the partition function of Eq. (\ref{partfun}), the 
constraint will be spurious;
in any approximate treatment of the functional integral
for $Z$, however,
(as for example in a coarse--graining approach) the introduction of
the unrestricted order parameter field generally will lead to a
violation of Eq.(\ref{spin2}). In such a case the introduction of the
spherical constraint on the order parameter field
 in the partition function simply reflects the constrained nature of the
original pseudo--spin variables.

Integrating over the phase variables leaves the statistical
sum entirely written in terms of  the constrained continuous order
parameter field   ${\mbox{\boldmath$\psi$}}_{\bf r}(\tau)$,
so that the partition function becomes
\begin{eqnarray}
Z&=&\int\left[\prod_{\bf r}
 {\cal D}{\mbox{\boldmath$\psi$}}_{\bf r}
{\cal D}{\mbox{\boldmath$\psi$}}^\star_{\bf r}\right]
\delta\left(\sum_{\bf r} |{\mbox{\boldmath$\psi$}}_{\bf r}(\tau)|^2
-N\right)
e^{-{\cal S}_{\rm eff}[{\mbox{\boldmath$\psi$}}]},
\label{statsum2}
\end{eqnarray}
where the effective order parameter action reads
\begin{equation}
{\cal S}_{\rm eff}[{\mbox{\boldmath$\psi$}}]=
{\cal S}_{J}[{\mbox{\boldmath$\psi$}}]+
\Phi_C[{\mbox{\boldmath$\psi$}}],
\end{equation}
with
\begin{equation}
{\cal S}_{J}[{\mbox{\boldmath$\psi$}}]=
\int_0^\beta d\tau
\sum_{\langle {\bf r}_1,{\bf r}_2\rangle}
J(|{\bf r}_1-{\bf r }_2|)
\psi_{{\bf r}_1}^\star(\tau)
\psi_{{\bf r}_2}(\tau),
\end{equation}
and
\begin{equation}
e^{-\Phi_C[{\mbox{\boldmath$\psi$}}]}=
\int\left[\prod_{\bf r}\frac{{\cal D}
  \mbox{\boldmath$\mu$}_{\bf r}}{2\pi i}\right]
\exp\left[-\left({\sum_{\bf r}\int_0^\beta
 d\tau\mbox{\boldmath$\mu$}_{\bf r}(\tau)
\cdot{\mbox{\boldmath$\psi$}}_{\bf r}(\tau)
-W[{\mbox{\boldmath$\mu$}}]}\right)\right],
\label{Phi}
\end{equation}
where the generating functional of the cumulant multipoint
phase correlators of the ``non--interacting" system
(i.e., involving only charging energy terms) is
\begin{equation}
W[{\mbox{\boldmath$\mu$}}]=\ln
\int\left[\prod_{\bf r} {\cal D}{\phi}_{\bf r}\right]
\exp\left({\sum_{\bf r}\int_0^\beta
 d\tau\mbox{\boldmath$\mu$}_{i}(\tau)
\cdot{\bf S}_{\bf r}(\phi(\tau))
+{\cal S}_C[{\phi}]}\right)
\label{W}
\end{equation}
 with the
variables ${\mbox{\boldmath$\mu$}}$ acting as the
source fields.
It can be seen that from Eq.(\ref{W}) that
$e^{-\Phi_C[{\mbox{\boldmath$\psi$}}]}$ is just the
functional Fourier transform of 
$e^{-W[{\mbox{\boldmath$\mu$}}]}$.
The standard way to proceed  is
to calculate $\Phi_C[{\mbox{\boldmath$\psi$}}]$
using the saddle point method and a subsequent loop expansion
in terms of the powers of the order parameter 
using Eq. (\ref{W}).  The structure of this expansion is briefly 
described in the Appendix.

It is convenient to employ the functional Fourier
representation of the $\delta$--functional to resolve the
 spherical constraint in Eq.(\ref{statsum2}):
\begin{eqnarray}
 \delta\left[g(\tau)\right]
=\int_{-i\infty}^{+i\infty}\left[\frac{{\cal D}
 \xi(\tau)}{2\pi i}\right]
\times\exp\left[\int_0^\beta d\tau \xi(\tau)g(\tau)\right].
\label{fourier}
\end{eqnarray}
Accordingly, we can write
\begin{eqnarray}
 \delta\left(\sum_{\bf r}
 |{\mbox{\boldmath$\psi$}}_{\bf r}(\tau)|^2\right.&-&N\Bigg)
\equiv\int_{-i\infty}^{+i\infty}\left[\frac{{\cal D}
 \lambda(\tau)}{2\pi i}\right]
\times\nonumber\\
&\times&\exp\left[\int_0^\beta d\tau \lambda(\tau)
\left(\sum_{\bf r} |{\mbox{\boldmath$\psi$}}_{\bf r}(\tau)|^2
-N\right)\right].
\end{eqnarray}
We may factor the trace over each ${\mbox{\boldmath$\psi$}}_{\bf r}(\tau)$
at the expense of introducing a new integral over one component field
$\lambda(\tau)$. Expanding part of the effective action 
$\Phi_C[{\mbox{\boldmath$\psi$}}]$ to second order in
${\mbox{\boldmath$\psi$}}_{\bf r}(\tau)$ we get from Eq.(\ref{statsum2})
\begin{eqnarray}
Z_{\rm QSA}
=\int\left[\prod_{\bf r} {\cal D}{\mbox{\boldmath$\psi$}}_{\bf r}
{\cal D}{\mbox{\boldmath$\psi$}}^\star_{\bf r}\right]
\int\left[\frac{ {\cal D}\lambda}{2\pi i}\right]
e^{-{\cal S}_{\rm QSA}[
{\mbox{\boldmath$\psi$}},\lambda]},
\label{statsumqsa}
\end{eqnarray}
where
\begin{eqnarray}
{\cal S}_{\rm QSA}[
{\mbox{\boldmath$\psi$}},\lambda]
&&= \int_0^\beta d\tau d\tau'
\left\{\sum_{\langle{\bf r}_1,{\bf r}_2\rangle}
\left[\left( {\bf J}(|{\bf r}_1-{\bf r}_2|
+\lambda(\tau)\delta_{{\bf r}_1,{\bf r}_2}\right)
\delta(\tau-\tau')\right.\right.
\nonumber\\
&&+
\left. \left.\Gamma_{02}^{-+}({\bf r}_1\tau;{\bf r}_2\tau')\right]
{\psi}^{\star}_{{\bf r}_1}(\tau)
{\psi}_{{\bf r}_2}(\tau')
-N\lambda(\tau) \delta(\tau-\tau')\right\},
\label{qsa}
\end{eqnarray}
is the effective action of the quantum spherical
approximation (QSA).
Subsequent improvements over the spherical model approach may be obtained 
by considering  the saddle point corrections to the Lagrangian  parameter
$\lambda$. This can be conveniently done in terms of a loop expansion as 
described in the Appendix.
%
Furthermore,
$\Gamma_{02}^{-+}({\bf r}_1\tau;{\bf r}_2\tau')
=[W_{02}^{-1}]^{-+}({\bf r}_1\tau;{\bf r}_2\tau')$
is the two--point phase vertex correlator and
\begin{equation}
W_{02}^{-+}({\bf r}_1\tau;{\bf r}_2\tau')=
\frac{1}{Z_0}\sum_{\{n_{\bf r}\}}\prod_{\bf r}\int_0^{2\pi}
 d\theta(0)\int_{\theta(0)}^{\theta(0)+2\pi n_{\bf r}}
{\cal D}\theta_{\bf r}(\tau)
e^{i[\theta_{{\bf r}_1}(\tau)-\theta_{{\bf r}_2}(\tau')]}
e^{-{\cal S}_C[\theta]},
\label{w02}
\end{equation}
with
\begin{equation}
Z_0=\sum_{\{n_{\bf r}\}}\prod_{\bf r}\int_0^{2\pi}
 d\theta(0)\int_{\theta(0)}^{\theta(0)+2\pi n_{\bf r}}
{\cal D}\theta_{\bf r}(\tau)
e^{-{\cal S}_C[\theta]},
\end{equation}
is  the partition function for the ``non--interacting" system.
Eq. (\ref{qsa})  incorporates the quadratic term
proportional to $\lambda(\tau)$ and the integration in
Eq.(\ref{statsumqsa}) takes place over all configurations
of the order parameter field. The model is now unconstrained
and quadratic, so all quantities can be readily computed.
In the thermodynamic limit, $N\to\infty$, we can
calculate the functional integral in Eq.(\ref{statsumqsa})
by the steepest--descent method. To proceed, we introduce propagators
associated with the order parameter field defined by
\begin{equation}
G({\bf r}_1\tau;{\bf r}_2\tau')={\langle {\psi}^{\star}_{{\bf r}_1}(\tau)
{\psi}_{{\bf r}_2}(\tau')\rangle}_{\rm QSA}.
\label{opcor}
\end{equation}
The condition that the integrand in Eq.(\ref{statsumqsa})
 has a saddle point $\lambda(\tau)=\lambda_0$ is that
\begin{equation}
1=\frac{1}{N}\sum_{\bf r}
G({\bf r}\tau;{\bf r}\tau+0^+),
\label{gconstr}
\end{equation}
which becomes an implicit equation for $\lambda_0$.
The QSA ensemble average is now defined by
\begin{equation}
 \langle \dots \rangle_{\rm QSA} =
\frac{\int \left[\prod_{\bf r} {\cal D}{\mbox{\boldmath$\psi$}}_{\bf r}
{\cal D}{\mbox{\boldmath$\psi$}}^\star_{\bf r}\right] \dots
e^{  -{\cal S}_{\rm QSA}[
{\mbox{\boldmath$\psi$}},\lambda_0]   }}
{\int\left[\prod_{\bf r} {\cal D}{\mbox{\boldmath$\psi$}}_{\bf r}
{\cal D}{\mbox{\boldmath$\psi$}}^\star_{\bf r}\right]
e^{  -{\cal S}_{\rm QSA}[
{\mbox{\boldmath$\psi$}},\lambda_0] }}.
\end{equation}
A Fourier transform of Eq. (\ref{qsa})
in momentum and frequency space enables one to write
the spherical constraint (\ref{gconstr}) explicitly as
\begin{equation}
1=\frac{1}{N\beta}\sum_{\bf k}\sum_\ell
\frac{1}{\lambda-J({\bf k})+\left[W_{02}(\omega_\ell)\right]^{-1}},
\end{equation}
with $J({\bf k})$ the Fourier transform of
the Josephson interactions ${\bf J}(|{\bf r}_1-{\bf r}_2|)$,
$W_{02}(\omega_\ell)$ the frequency transformed
phase--phase correlator of Eq. (\ref{w02})
and $\omega_\ell=2\pi\ell/\beta$ $(\ell=0,\pm 1,\pm 2,\dots)$
the (Bose) Matsubara frequencies.
As usual in a spherical model analysis the critical behavior
is determined by the denominator of the summand in the
spherical constraint equation of Eq. (\ref{gconstr}).
Specifically, when [$1/G({\bf k}=0,\omega_\ell=0)]=0$,
where $G({\bf k},\omega_\ell)$ is the Fourier transformed
order parameter correlation of Eq. (\ref{opcor}), the system displays
a critical point at
\begin{equation}
\lambda_0-J({\bf k}=0)+\left[W_{02}(\omega_\ell=0)\right]^{-1}=0,
\label{critcond}
\end{equation}
provided the momentum  and frequency summations in the constraint
in Eq. (\ref{gconstr}) converge. The system
does not show a phase transition if the frequency--momentum
sum in Eq.(\ref{gconstr}) diverges for $N\to\infty$ at the
point defined by the criticality condition (\ref{critcond}).
The universal critical properties only depend on the
low--frequency behavior of $\left[W_{02}(\omega_\ell)\right]^{-1}$
and the long--wavelength properties of the interaction $J({\bf k})$.
Formally, the $T=0$ critical behavior will be identical to that of
interacting quantum rotors with a kinetic energy proportional
to $k^2$ in $d=3$ dimensions, i.e. the transition will be
in the universality class of the tree dimensional (3D)
XY model.

\subsection{Zero temperature ground capacitance model.}

Offset charges are an important ingredient in the experimental array
 samples made of ultrasmall
junctions. In this subsection we reconsider this problem as a test for
 the usefulness of the quantum spherical model
approach.
The offset charges, or an external gate voltage
applied between the array and the substrate, behave like a chemical
potential for injection of Cooper pairs into the
array. Several authors have shown that 
static background charges can have a pronounced
effect on the SI transition
at zero temperature\cite{Bruder,Freericks}. 
Including the offset charges, $q_x$,
in the charging energy of Eq. (\ref{action}) gives
\begin{equation}
{\cal S}_C[\phi]=\frac{1}{8e^2} \int_0^\beta d\tau
\sum_{\bf r,r'}
[{\bf C}^{-1}]_{{\bf r}{\bf r}'}
\left(\frac{\partial\phi_{\bf r} }{\partial\tau}
-q_x\right)
 \left(\frac{\partial\phi_{\bf r'} }{\partial\tau}
-q_x\right).
\end{equation}
Furthermore, we simplify the model to include only the 
background capacitance  (or self--charging) model.
In this case $[{\bf C}^{-1}]_{{\bf r},{\bf r}'}
=\delta_{{\bf r},{\bf r}'}/C_0$
and $E_C=\frac{1}{2}e^2[{\bf C}^{-1}]_{{\bf r},{\bf r}}\equiv e^2/2C_{0}$.
Of course the approximation that $C_0 >> C_1$, where $C_1$ is the mutual
capacitance between the islands,
is a first step in the analysis that leads to physical insights into the general problem.
Finally, we assume a square array characterized by the nearest--neighbor Josephson coupling $E_J$
 with $J({\bf k})=E_J[\cos(k_xa)+\cos(k_ya)]$,
 
with $a$ the lattice constant.
We obtain the corresponding density of states
\begin{equation}
\rho(E)=\frac{1}{N}\sum_{\bf k}{
\delta(E-J({\bf k}))}=\frac{1}{\pi^2E_J}
{\bf K}\left(\sqrt{1-\frac{E^2}{4E_J^2}}\right)\Theta(2E_J-|E|),
\label{density}
\end{equation}
where ${\bf K}(x)$ is the complete elliptic
integral of the first kind~\cite{stegun}.
The phase--phase correlation function becomes\cite{Bruder}:
\begin{eqnarray}
W_{02}(\omega_\ell)=
\frac{8E_C}{Z_0}\sum_{q}
\frac{\displaystyle\exp\left[-4\beta E_C
(q-q_x)^2\right]}
{\displaystyle(4E_C)^2-\left[8E_C(q-q_x)	
-i\omega_\ell\right]^2},
\label{w02q}
\end{eqnarray}
where $Z_0=\sum_{q}\exp\left[-4\beta E_C
(q-q_x)^2\right]$, and the summation is performed over all
integer--valued charge states  $q=0,\pm 1,\pm 2,\dots$, which makes
the function $W_{02}(\omega_\ell)$ periodic. At low temperatures
 the sum over $q$ in Eq.(\ref{w02q})
is dominated by the charge $q$ which makes the exponent in the
numerator of  Eq.(\ref{w02q}) smallest. For $T=0$
this value is $q=0$ and
the equation for the critical line ($E/E_C$) vs. $q_x$ 
is obtained from the implicit equation
\begin{equation}
1=\frac{1}{\beta}\sum_\ell\int^{+\infty}_{-\infty}dE
\frac{\rho(E)}{\lambda_0+2E_C-E- 8E_C\left(q_x+\frac{i\omega_\ell}{8E_C}
\right)^2 },
\label{2cond}
\end{equation}
valid for $-\frac{1}{2}\le q_x\le \frac{1}{2}$
(other values of $q_x$ are included by a periodic extension
with the period equal to one). The criticality condition
(cf. Eq.(\ref{critcond})) reads
\begin{equation}
\lambda_0=2E_J-2E_C+8E_cq_x^2.
\label{cr2} 
\end{equation}
By formally setting  $\lambda_0=0$ in Eq.(\ref{cr2})
we obtain the coarse-grained mean--field parabolic solution:
\begin{equation}
\frac{E_J}{E_C}=1-4q_x^2.
\label{naiveMF}
\end{equation}
By substituting the value of $\lambda_0$
from Eq.(\ref{cr2}) into (\ref{2cond}) within the QSA method, after performing the
summation over Matsubara
frequencies, one obtains the $T\to 0$ limit result
\begin{eqnarray}
1={\cal P}\int^{+\infty}_{-\infty}dE\rho(E)
\sqrt{\frac{E_C}{2(\lambda_0+2E_C-E)}}
&&\left[{\rm sign}\left(4q_x\frac{E_C}{E_J}
+\frac{\sqrt{2(\lambda_0+2E_C-E)E_C}}{E_J}\right)\right.
\nonumber\\
&&-\left.{\rm sign}\left(4q_x\frac{E_C}{E_J}
-\frac{\sqrt{2(\lambda_0+2E_C-E)E_C}}{E_J}\right)\right],
\end{eqnarray}
where ${\cal P}$ denotes the principal value of the integral.
After Eq.(\ref{density}) we finally obtain
the phase boundary of the insulating Mott lobe at
zero temperature
\begin{eqnarray}
1=\frac{\sqrt{2}}{\pi^2}\int^{2}_{-2}dy
\frac{\displaystyle{\bf K}\left(\sqrt{1-{y^2}/{4}}\right)}
{\displaystyle\sqrt{(2-y)\frac{E_J}{E_C}+8q_x^2}}.
\end{eqnarray}
In Fig.(\ref{fig1}) we plot the resulting phase boundary in the
($E_J/E_C$)--vs--$q_x$ plane at $T=0$. We recognize the
periodic lobes in $q_x$ of the insulating phase separated by regions
of phase coherent superconducting state. For $q_x=1/2$
the superconducting state extends down to arbitrarily small values
of $E_J/E_C$. We can compare the phase diagram resulting from the QSA to
the one obtained from mean--field theory via
the coarse grained approach (cf. Eq.(\ref{naiveMF})). Whereas
QSA gives the value $E_J(q_x=0)/E_C\approx 1.653$,
mean--field theory gives $E_J(q_x=0)/E_C=1$,
which underestimates the critical value of $E_J$.
It is also instructive to compare the QSA result for $E_J(q_x=0)/E_C$
to the recent third--order perturbation
expansion in $E_J/E_C$~\cite{kim}:
\begin{equation}
q_x=\frac{1}{2}-\frac{1}{4}\frac{E_J}{E_C}-\frac{3}{128}
\left( \frac{E_J}{E_C} \right)^2-\frac{11}{1025}\left( \frac{E_J}{E_C} \right)^3
+O\left[\left( \frac{E_J}{E_C} \right)^4\right],
\end{equation}
which gives $E_J(q_x=0)/E_C\approx 1.590$, giving a $3.8\%$ of the
difference from the critical value of $E_J/E_C$ at $q_x=0$
obtained in the QSA.

\section{Conductivity scaling}

The conductivity is an experimentally measurable quantity in JJA.
There are other studies of  
the superconductor--Mott--insulator and its universal conductance in JJA.
Cha et al (see, Ref.\onlinecite{cha})
carried out a $1/N$ expansion and Monte Carlo analysis.
Van Otterlo et. al. used the coarse--graining approach\cite{otterlo}
and Fazio and Zappela did an $\epsilon$-expansion\cite{fazio}.
Here, we are interested in another aspect of
the SI transition, namely, the
scaling of the conductivity in the vicinity of
the quantum critical point.

To evaluate the conductivity, we need to add an external field in terms of 
a minimally coupled order parameter to the
vector potential ${\bf A}({\bf r},\tau)$. The Josephson coupling term
in Eq. (\ref{action}) then becomes
\begin{equation}
J(|{\bf r}_i-{\bf r}_j|)\to J(|{\bf r}_i-{\bf r}_j|)
\exp\left(i\frac{2e}{\hbar}\int_{{\bf r}_i}^{{\bf r}_j}
{\bf A}\cdot d{\bf l}  \right).
\end{equation}
The phase shift on each junction 
is determined by the vector potential ${\bf A}$ of the magnetic applied field
 and in a typical experimental situation it
can be entirely ascribed to the external field.

The imaginary--time frequency dependent linear-response conductivity is given by
\begin{equation}
\sigma_{\mu\nu}(\omega_\nu,{\bf q})=\frac{\hbar}{\omega_\nu}
\int d^2{\bf r}\int_0^\beta d\tau\frac{\delta^2\ln Z}
{\delta A_\mu(\tau,{\bf r})\delta A_\nu(0,0)}
e^{i\omega_\nu\tau+i{\bf q}{\bf r}}.
\end{equation}
For vanishing magnetic field and offset charge the longitudinal
component of $\sigma(\omega_\nu)\equiv\sigma_{xx}(\omega_\nu,{\bf q}=0)$
 in the QSA is
\begin{equation}
\sigma(\omega_\nu)=\frac{4e^2E_J^2}{\hbar\omega_\nu}
\int_{-\pi/a}^{\pi/a}\frac{dk_xdk_y}{(2\pi/a)^2}
\frac{1}{\beta}\sum_{\omega_\ell}\sin^2(ak_x)\left[
G^2({\bf k},\omega_\ell)-
G({\bf k},\omega_\ell)G({\bf k},\omega_\nu+\omega_\ell)
\right],
\label{conduct}
\end{equation}
where $G^{-1}({\bf k},\omega_\ell)=[{\lambda-J({\bf k})+2E_C+
{\omega_\ell^2}/{8E_C}}]$, and $\lambda$ is determined
self--consistently from the constraint equation
(\ref{constr}). To proceed it is convenient to obtain
the generalized density of states for the 2-D square lattice as
\begin{eqnarray}
&&\rho_2(E)=\int_{-\pi/a}^{\pi/a}\frac{dk_xdk_y}{(2\pi/a)^2}
\sin^2(ak_x)\delta(E-J({\bf k}))=
\nonumber\\
&&=
\frac{2}{\pi^2E_J}\left[{\bf E}\left(\sqrt{1-\frac{E^2}{4E_J^2} }\right)
-\left( \frac{E}{2E_J} \right)^2{\bf K}\left(\sqrt{1-\frac{E^2}{4E_J^2} }
\right)
\right]\Theta(4E_J^2-E^2)
\label{rho2}
\end{eqnarray}
where ${\bf E}(x)$ is the elliptic integral of the second kind.
In terms of Eq. (\ref{rho2}) the conductivity of Eq. (\ref{conduct}) becomes
\begin{equation}
\sigma(\omega_\nu)=\frac{4e^2E_J^2}{\hbar\omega_\nu}
\int_{-\infty}^{+\infty} dE
\frac{1}{\beta}\sum_{\omega_\ell}\rho_2(E)G(E,\omega_\ell)\left[
G(E,\omega_\ell)-G(E,\omega_\nu+\omega_\ell)
\right].
\label{conduct2}
\end{equation}
Here,
$G^{-1}(E,\omega_\ell)=[{\delta_\lambda-E+2E_J+
{\omega_\ell^2}/{8E_C}}]$,
where $\delta_\lambda=\lambda-\lambda_{\rm crit}$.
The zero--temperature critical boundary of the phase  coherent
state is signaled by $\lambda(E_J/E_C,T)$ reaching the
value $\lambda_{\rm crit}(E^{\rm crit}_J/E_C,T=0)$.
The parameter $\delta_\lambda$ plays the important role of
energy scale (which vanishes in the
superconducting phase). We now consider the $T=0$ phase transition
between the long--range ordered coherent phase  state and
the insulating phase by varying the coupling constant $E_J$
through the critical value $E_J^{\rm crit}$, where there is
a diverging correlation length $\xi\sim|E_J-E_J^{\rm crit}|$.
At finite temperatures, the deviations from
$T=0$ behavior are distinguished by the scale set by the thermal coherence length
$\xi_T\sim T^{-1/z}$ (where $z$ is the dynamic critical
exponent). The quantum--critical region
is defined by the inequality $\xi_T<\xi$.
In this case the system feels the finite value of the temperature
before becoming sensitive to the deviations of $E_J$ from
$E_J^{\rm crit}$. In this regime the dynamic conductivity
turns out to be remarkably universal.
To proceed, we need to determine the $\delta_\lambda$ dependence
of temperature and $E_J/E_C$.
The spherical constraint of Eq. (\ref{constr}) takes the form
\begin{eqnarray}
1=\frac{1}{\beta}\sum_\ell\int^{+\infty}_{-\infty}dE
\frac{\rho(E)}{\delta_\lambda+2E_J-E
- 8E_C\left(\frac{i\omega_\ell}{8E_C}
\right)^2 }.
\end{eqnarray}
The near--critical properties of the spherical model are
 essentially determined by the
structure of the spectrum of the interaction matrix $J(|{\bf r}_1-{\bf r}_2|)$
in the neighborhood of its upper boundary, specifically by the
behavior of the density of states associated with $J(|{\bf r}_1-{\bf r}_2|)$.
We therefore expand the square lattice density of states (\ref{density}) 
about the upper limit of the spectrum of $J({\bf k})$:
\begin{equation}
\frac{1}{{{E_J}{\pi }^2}}{{\bf K}\left[\sqrt{1 - 
       {\frac{{{\left(E -2\, {E_J}  \right) }^2}}
         {4\,{{ {E_J}}^2}}}}\right]} = 
  {\frac{1}{2\, {E_J}\,\pi }} + 
   {\frac{E}{8\,{{ {E_J}}^2}\,\pi }} +    
   {{{O}(E^2)}}.
\end{equation}
The subsequent summation frequency yields
\begin{equation}
1=\frac{1}{2\pi}\frac{k_BT}{E_J}\ln
\left\{
\frac{
\sinh\left[\beta\sqrt{2E_C(4E_J+\delta_\lambda})\right]}
{
\sinh\left[\beta\sqrt{2E_C(\delta_\lambda})\right]}
\right\}.
\end{equation}
This constraint equation can be explicitly solved for the
dependence of $\delta_\lambda$ on $T,E_J,$ and $E_C$ giving
\begin{equation}
\delta_\lambda=\frac{(k_BT)^2}{2E_c}\left\{{\rm arsinh}\left[e^{-2\pi\beta E_J}
 \sinh \left(\beta\sqrt{8E_CE_J} \right)\right]\right\}^2,
\end{equation}
and near the $T=0$ quantum critical point
\begin{equation}
\delta_\lambda=\frac{(k_BT)^2}{2E_C}\left\{{\rm arsinh}\left[
 \frac{1}{2}\exp \left(-\pi\beta(E_J-E_J^{\rm crit})
 \right)\right]\right\}^2,
 \label{solconstr}
\end{equation}
where ${E_J^{\rm crit}}={2E_C}/{\pi^2}$.
We extract the conductivity, as a function of the
real frequencies $\omega$ in the form of real and imaginary parts
of $\sigma(\omega)\equiv\sigma_{\rm sing}(\omega_\nu)
+\sigma_{\rm reg}(\omega_\nu)$, after performing the sum over Matsubara frequencies in Eq.(\ref{conduct2})
followed by analytic continuation to real frequencies
$i\omega_\ell\to\omega +i0^+$ 
getting
\begin{eqnarray}
&&{\rm Im}\sigma_{\rm sing}(\omega)=
\frac{4e^2E_J^2}{\hbar}
\int_{-\infty}^{+\infty} dE
\rho_2(E)
\frac{1}{\omega}\frac{8\beta E_C {\rm cosech}^2\left(\frac{\beta}{2}
\sqrt{8E_C(\delta_\lambda-E+2E_J})\right)}{\delta_\lambda-E+2E_J},
\nonumber\\
&&{\rm Im}\sigma_{\rm reg}(\omega)=
\frac{4e^2E_J^2}{\hbar}
\int_{-\infty}^{+\infty} dE
\rho_2(E)\frac{1}{4}\sqrt{\frac{1}{8E_C(\delta_\lambda-E+2E_J)^3}  }
\frac{\omega\coth\left(\frac{\beta}{2}
\sqrt{8E_C(\delta_\lambda-E+2E_J})\right)}
{\frac{\omega^2}{8E_C}-4(\delta_\lambda-E+2E_J  ) }
\label{scal1cond}
\end{eqnarray}
and
\begin{eqnarray}
&&{\rm Re}\sigma_{\rm sing}(\omega)=
\frac{4e^2E_J^2\pi}{\hbar}
\int_{-\infty}^{+\infty} dE
\rho_2(E)
\delta({\omega})\frac{8\beta E_C {\rm cosech}^2\left(\frac{\beta}{2}
\sqrt{8E_C(\delta_\lambda-E+2E_J})\right)}{\delta_\lambda-E+2E_J}
\nonumber\\
&&{\rm Re}\sigma_{\rm reg}(\omega)=
\frac{2\cdot 4e^2E_J^2\pi}{\hbar}
\int_{-\infty}^{+\infty} dE
\rho_2(E)\frac{1}{4}\frac{1}{(\delta_\lambda-E+2E_J)}  
{\coth\left(\frac{\beta}{2}
\sqrt{8E_C(\delta_\lambda-E+2E_J})\right)}
\nonumber\\
&&\times
\delta\left[\frac{\omega^2}{8E_C}-4(\delta_\lambda-E+2E_J)  \right].
\label{condu3}
\end{eqnarray}
The real part of the conductivity contains two contributions:
the first, ${\rm Re}\sigma_{\rm sing}(\omega)$, is singular since it is proportional to
$\delta(\omega)$ and the second finite--frequency is regular, 
 ${\rm Re}\sigma_{\rm reg}(\omega)$, which
arises from the electromagnetic field induced
transitions to excited states. The singular part in turn is due to
the free acceleration of quasiparticles. This is so since the
JJA model considered here contains no dissipation mechanism which would e.g.
arise from a coupling of the phase degrees of freedom
to the normal electrons (Ohmic damping). At $T=0$ the singular part vanishes
while the regular one can be evaluated explicitly by performing an
energy integration in Eq.(\ref{condu3}) with the result
\begin{eqnarray}
{\rm Re}\sigma_{\rm reg}(\omega,\delta)=\frac{(2e)^2}{h}\frac{1}{2 {\delta} }
\left(\frac{\omega_c}{ \omega }\right)^2
{\cal F}\left[1+\frac{ {\delta} }{2}
\left(1-\frac{{\omega}^2}{\omega_c^2}\right) \right]
\Theta\left[\left(\left|\frac{\omega}{\omega_c}\right|-1\right)
\left(\sqrt{1+4/{\delta}}
-\left|\frac{\omega}{\omega_c}\right|\right)  \right],
\label{qsacond}
\end{eqnarray}
where ${\cal F}(x)={\bf E}\left(\sqrt{1-x^2} \right)
-x^2{\bf K}\left( \sqrt{1-x^2} \right)$,
$\bar{\omega}=\omega/\omega_c$, 
with $\omega_c=\sqrt{32E_C\delta_\lambda}$
and $\delta=\delta_\lambda/E_J$.
Here $\omega_c$ denotes the thrashed frequency above which the 
particle--hole excitations can be created and the real part
of the conductivity is finite, while ${\rm Re}\sigma_{\rm reg}(\omega)$
vanishes for $\omega<\omega_c$ indicating a Mott--insulating phase.
Letting the gap parameter $\delta$ go to zero, while keeping
the ratio $\omega/\omega_c$ finite, and using the result
\begin{equation}
\lim_{x\to 0}
\frac{2}{{{x\pi }^2}}   {\cal F}(1-x) 
       = {\frac{1}{\pi }},
\end{equation}
we obtain from the general QSA result (\ref{qsacond})
(valid for arbitrary distance $\delta_\lambda$ away
from the critical point) the long--wave length limit near--critical  form
of the conductivity:
\begin{equation}
\lim_{\delta\to 0}{\rm Re}\sigma_{\rm reg}
(\omega,\delta)=\frac{(2e)^2}{h}\frac{\pi}{8 }
\left(1-\frac{\omega_c^2}{\omega^2}     \right)
\Theta\left(\left|\frac{\omega}{\omega_c}\right|-1\right).
\label{longcond}
\end{equation}
A plot of this result for different values of $\delta$ is shown in 
Fig. (\ref{fig2}), where we clearly see an asymptotic gap as 
we approach the critical point at zero temperature.
This analytical result was previously derived in studies 
that relied on the coarse--grained
and loop--expansion approaches. From Eq.(\ref{longcond})
at the transition, where the thrashed
frequent $\omega_c$ vanishes,  a finite dc conductance
$\sigma^\star=(\pi/8)\cdot 4e^2/h$ emerges which is the universal
zero temperature conductivity earlier obtained by Cha {\it et al}.
In practice, however, all experiments are performed
at low but nonzero temperature. 

We will now present the scaling analysis satisfied  by
$\sigma(\omega)$ in the vicinity of the quantum phase transition
$E_J=E_J^{\rm crit}$. The temperature is taken to be nonzero but
must obey $k_BT<<E_J$. A nonzero $T$ implies the absence of
long--range phase coherence and the scaling properties will
depend upon a variable which measures the distance of the
superconducting ground state from criticality. The
behavior of the conductivity in this regime can be
understood in terms of the scaling forms
\begin{eqnarray}
&&{\rm Re}\sigma(\omega)=
\frac{(2e)^2}{h}g'\left(\frac{\omega}{k_BT},
\frac{(k_BT)^{1/\nu z}}{\delta E_J}   \right),
\nonumber\\
&&{\rm Im}\sigma(\omega)=
\frac{(2e)^2}{h}g''\left(\frac{\omega}{k_BT},
\frac{(k_BT)^{1/\nu z}}{\delta E_J}   \right),
\end{eqnarray}
where ${g'}(X,Y)={g'}_{\rm sing}(X,Y)+{g'}_{\rm reg}(X,Y)$
and ${g''}(X,Y)={g''}_{\rm sing}(X,Y)+{g''}_{\rm reg}(X,Y)$
are highly non--trivial but universal two--parameter functions.
>From equations (\ref{scal1cond}) and (\ref{condu3})
and the explicit solution (\ref{solconstr}) of the spherical  constraint
equation we derive
\begin{eqnarray}
{g'}_{\rm sing}(X,Y)&=&
\frac{\pi}{2}\delta(X){\Xi^2(Y)}\int_1^\infty du\left(u-\frac{1}{u}\right)
\Sigma(Y,u),
\nonumber\\
{g'}_{\rm reg}(X,Y)&=&\frac{\pi}{8}\left[
1-\frac{\Xi^2(Y)}{X^2}
\right]\coth\left(\frac{X}{4}  \right)
\Theta\left[\frac{X^2}{\Xi^2(Y)} -1\right],
\end{eqnarray}
where
\begin{eqnarray}
&&\Xi(Y)=4\ln\left(\frac{1}{2}\sqrt{e^{-2\pi/Y}+4}+\frac{1}{2}e^{-\pi/Y}
  \right)\nonumber\\
&&\Sigma(Y,u)=\frac{ 4 }  {\left[\left(\frac{1}{2}
\sqrt{e^{-2\pi/Y}+4}+\frac{1}{2}e^{-\pi/Y}
  \right)^u- \left(\frac{1}{2}
\sqrt{e^{-2\pi/Y}+4}+\frac{1}{2}e^{-\pi/Y}
  \right)^{-u} \right]^2 },
\end{eqnarray}
and
\begin{eqnarray}
{g''}_{\rm sing}(X,Y)&=&\frac{\pi}{2}\frac{\Xi^2(Y)}{X
}\int_1^\infty du\left(u-\frac{1}{u}\right)
\Sigma(Y,u)
\nonumber\\
{g''}_{\rm reg}(X,Y)&=& \frac{1}{4}\frac{X}{\Xi(Y)}
\int_{1}
^{\infty}du\left(1-\frac{1}
{u^2}\right)\frac{\Omega(Y,u)}{\frac{X^2}{\Xi^2(Y)}-u^2}
\end{eqnarray}
with
\begin{equation}
\Omega(Y,u)=\frac{ \left(\frac{1}{2}
\sqrt{e^{-2\pi/Y}+4}+\frac{1}{2}e^{-\pi/Y}
  \right)^u+ \left(\frac{1}{2}
\sqrt{e^{-2\pi/Y}+4}+\frac{1}{2}e^{-\pi/Y}
  \right)^{-u} }{\left(\frac{1}{2}
\sqrt{e^{-2\pi/Y}+4}+\frac{1}{2}e^{-\pi/Y}
  \right)^u- \left(\frac{1}{2}
\sqrt{e^{-2\pi/Y}+4}+\frac{1}{2}e^{-\pi/Y}
  \right)^{-u}  }.
\end{equation}
We are in the quantum--critical regime by 
setting $\delta E_J$ to zero  ($Y=\infty$), while keeping the
temperature small but finite ($X\neq 0$),
with $z=2$, and $z\nu=1$, with the critical
fluctuations quenched in a universal way by temperature.
Here, $k_BT$ appears to be the dominant energy
which determines the physics of the problem. At large frequencies,
$\hbar\omega>>k_BT$, the finite temperature effects
do not become manifest and the system displays
the behavior of the zero--temperature critical point
($\delta E_J=0, T=0$). However, at small enough frequencies,
$\hbar\omega\sim k_BT$, the finite temperature effects of a
become apparent which introduces a new energy scale into the problem.
While fluctuations with frequencies larger then $k_BT/\hbar$
are unaffected those with $\hbar\omega<k_BT$ will behave
classically. We show the regular parts of the nontrivial scaling functions
$g'(X,Y)$ and $g"(X,Y)$ in Figs. (\ref{fig3}) and (\ref{fig4}).

\section{Summary}
In this paper we have developed a microscopic analysis of a coupled 
array of ultrasmall Josephson junctions in the presence of 
charging energy effects. Our analysis is based on the quantum 
spherical model approach and the path--integral formulation
of quantum mechanics explicitly tailored for the microscopic
JJA Hamiltonian. The effective action formalism allows for 
an explicit implementation of the Coulomb  and offset voltage 
effects into our consideration. Using this formalism we have considered
the zero--temperature phase transition to try to understand the 
finite temperature behavior of the system in terms of
scaling functions for the electromagnetic response of the array. 
This is important since the experimental analysis of quantum 
phase transitions relies on the scaling behavior
of observables (eg. the conductivity) for relevant
physical parameters  (like temperature and frequency).
A relevant question which naturally arises is: To what extent 
does  this phenomena depend on the model used to describe the JJA?
Future theoretical studies on quantum critical behavior
should consider disorder effects (random offset charges), coupling
to quasiparticles (dissipation), magnetic field frustration and 
finite size effects, just to call a few relevant issues in this problem. 
The formalism presented in this paper can be used to look at 
these specific situations.

\acknowledgments
This work has been partially supported  by  a NATO Collaborative Research Grant
No. OUTR.CRG 970299, by the Polish Science Committee (KBN)
under the grant No. 2P03B--02415 and by the National Science Foundation
grant DMR-9521845.

\newpage

\appendix
\section{Loop expansion for the effective action}

The generating functional of Eq.(\ref{W}) for the connected cumulant functions
$W_{0m}(x_1,\dots ,x_m)$ is given by the functional expansion
\begin{eqnarray}
W[{\mbox{\boldmath$\mu$}}]=
\sum_{m=1}^{\infty}\frac 1{m!}
\int dx_1
\dots dx_{m}
&&W_{0m}(x_1,\dots ,x_m)
\mu(x_1)\dots \mu(x_m),
\label{w2}
\end{eqnarray}
where we have for convenience introduced  the short--hand notation
\begin{eqnarray}
x_m&\equiv& ({\bf r}_m,\tau_m,a_m)
\nonumber\\
\int dx\dots&\equiv&
\int_0^\beta d\tau_m\sum_{{\bf r}_m}\sum_{a_m}\dots,
\end{eqnarray}
to obtain
\begin{equation}
W_{0m}(x_1,\dots ,x_m)
=\langle S^{a_1}_{{\bf r}_1}[\phi(\tau
_1)]\dots S^{a_m}_{{\bf r}_m}[\phi(\tau_m) ]\rangle _0^{{\rm cum}}.
\end{equation}
For a
given set of operators ${\cal A}$,${\cal B}$ and ${\cal C}$ the cumulant averages
 are defined as: $
\langle {\cal A}{\cal B}\rangle _0^{\rm cum}
=\langle {\cal A}{\cal B}\rangle _0
-\langle {\cal A}\rangle _0\langle
{\cal B}\rangle _0$; $\langle {\cal A}{\cal B}{\cal C}\rangle _0^c=
\langle {\cal A}{\cal B}{\cal C}\rangle _0-\langle
{\cal A}\rangle _0\langle {\cal B}{\cal C}\rangle _0-\langle {\cal B}
\rangle _0\langle {\cal A}{\cal C}\rangle
_0-\langle {\cal C}\rangle _0\langle{\cal  A}{\cal B}\rangle _0+2\langle
 {\cal A}\rangle _0\langle
{\cal B}\rangle _0\langle {\cal C}\rangle _0$, $\dots $ etc...
 where
 \begin{equation}
 \langle \dots \rangle_0 =
\frac{\int\left[\prod_{\bf r} {\cal D}{\phi}_{\bf r} \right]\dots
e^{  -{\cal S}_C[{\phi}] }}
{\int\left[\prod_{\bf r} {\cal D}{\phi}_{\bf r} \right]
e^{  -{\cal S}_C[{\phi}] } },
\end{equation}
which is closely related to the multi--point vertex function
$\Gamma_{0m}(x_1,\dots ,x_m)$. 
The generating functional reads:
\begin{eqnarray}
\Gamma[{\mbox{\boldmath$\psi$}}]=
\sum_{m=1}^{\infty}\frac 1{m!}
\int dx_1
\dots
 dx_{m}
&&\Gamma_{0m}(x_1,\dots ,x_m)
\psi(x_1)\dots \psi(x_m).
\label{gama}
\end{eqnarray}
The relation between $W[{\mbox{\boldmath$\mu$}}]$
and $\Gamma[{\mbox{\boldmath$\psi$}}]$ is then given by
the  Legendre transform 
\begin{equation}
\label{legendre}
\Gamma[{\mbox{\boldmath$\psi$}}]
+W[{\mbox{\boldmath$\mu$}}]
=\int dx\mu(x)
\psi(x).
\end{equation}
\widetext
We may now evaluate the action $\Phi_C[{\mbox{\boldmath$\psi$}}]$
defined in Eq.(\ref{Phi}), by introducing in the exponential of Eq.(\ref{Phi})
a formal parameter $\zeta$ to order the perturbation expansion
($\zeta$ will be set to unity at the end of calculation). We obtain
\begin{equation}
e^{-\Phi_C[{\mbox{\boldmath$\psi$}}]}=
\int\left[\prod_{\bf r}\frac{{\cal D}
  \mbox{\boldmath$\mu$}_{\bf r}}{2\pi i}\right]
\exp\left[-\zeta\left({\sum_{\bf r}\int_0^\beta
 d\tau\mbox{\boldmath$\mu$}_{\bf r}(\tau)
\cdot{\mbox{\boldmath$\psi$}}_{\bf r}(\tau)
-W[{\mbox{\boldmath$\mu$}}]}\right)\right].
\label{PhiC}
\end{equation}
The steepest descent procedure is the standard method to
obtain the expansion for $\Phi_C[{\mbox{\boldmath$\psi$}}]$.
The first step is to
determine the saddle point ${\mbox{\boldmath$\mu$}}_{0}$
of the integrand in Eq.(\ref{PhiC}) followed by a systematic expansion
in terms of fluctuations about the saddle point. The procedure
is by now standard (see, e.g. Ref.\onlinecite{zinn}) and we quote the final result:
\begin{eqnarray}
&&e^{-\Phi_C[{\mbox{\boldmath$\psi$}}]}=
\exp\left[-\zeta\left(\int
 dx\mu_0(x)\psi(x)
-W[{\mbox{\boldmath$\mu$}}_{0}]\right)-
\frac{1}{2}\mbox{\rm tr}\ln W^{(2)}({\mbox{\boldmath$\psi$}})\right]\times
\nonumber\\
&&\times \left\{1+\frac{1}{\zeta}\left[-\frac{1}{8}
\int dx_1\dots dx_4 W^{(4)}(x_1,\dots,x_4;{\mbox{\boldmath$\psi$}})
D(x_1,x_2;{\mbox{\boldmath$\psi$}})D(x_3,x_4;
{\mbox{\boldmath$\psi$}})\right.\right.
\nonumber\\
&&+\left.\left.\int dx_1\dots dy_1
W^{(3)}(x_1,x_2,x_3;{\mbox{\boldmath$\psi$}})
W^{(3)}(y_1,y_2,y_3;{\mbox{\boldmath$\psi$}})
\left(\frac{1}{8}D(x_1,x_2;{\mbox{\boldmath$\psi$}})
\right.\right.\right.\times
\nonumber\\
&&\times\left.\left.\left.
D(y_1,y_2;{\mbox{\boldmath$\psi$} })
D(x_3,y_3;{\mbox{\boldmath$\psi$} })+\frac{1}{12}
D(x_1,y_1;{\mbox{\boldmath$\psi$} })
D(x_2,y_2;{\mbox{\boldmath$\psi$} })
D(x_3,y_3;{\mbox{\boldmath$\psi$}})\right)
\right] +O(\zeta^{-2})\right\},
\label{expans}
\end{eqnarray}
where
\begin{eqnarray}
\label{saddle}
&&\psi(x )=\left.\frac{\delta
 W[{\mbox{\boldmath$\mu$}}]}{\delta
\mu(x )} \right|_{\mbox{\boldmath$\mu$}=
{\mbox{\boldmath$\mu$}}_0(x;{\mbox{\boldmath$\psi$}})}
+\frac{1}{2\zeta}\int dy_1 dy_2
W^{(3)}(x,y_1,y_2;{\mbox{\boldmath$\psi$}})D(y_1,y_2;
{\mbox{\boldmath$\psi$}})+O(\zeta^{-2}),
\nonumber\\
&&{\mbox{\boldmath$\mu$}}_0(x;{\mbox{\boldmath$\psi$}})=\psi(x)
+\frac{1}{2\zeta}\int dy\ dy_1 dy_2
W^{(3)}(x,y_1,y_2;{\mbox{\boldmath$\psi$}})D(x,y;
{\mbox{\boldmath$\psi$}})D(y_1,y_2;{\mbox{\boldmath$\psi$}})
+O(\zeta^{-2}).
\end{eqnarray}
Here the propagator $D(x_1,x;{\mbox{\boldmath$\psi$}})$
is defined by
\begin{equation}
\int dx\zeta D(x_1,x;{\mbox{\boldmath$\psi$}})
W^{(2)}(x,x_2)=\delta(x_1-x_2),
\end{equation}
while
\begin{equation}
W^{(m)}(x_1,\dots,x_m)=
\frac{\delta^m W}{\delta \mu(x_1)\dots\delta\mu(x_m)},
\end{equation}
(cf. definition (\ref{w2})). Restricting ourselves to the lowest
order in the expansion (\ref{expans}) (only including the saddle point
terms proportional to $\zeta$) we obtain the
form of the effective action used in the QSA approach
(see, Eq.(\ref{qsa})).    

\begin{figure}
\caption{Zero--temperature phase phase boundary in the
$E_J/E_C$--vs--$q_x$  plane separating the Mott--insulating 
and superconducting states obtained from the 
quantum spherical approach (solid line), and the 
coarse--graining method (broken line). Note that there is no cusp
at $q_x=0$ since $dE_J(q_x)/dq_x|_{q_x=0}=0$ in the QSA.}
\label{fig1}
\end{figure}

\begin{figure}
\caption{Regular contribution of the
 frequency dependent conductivity at $T=0$ (real part)
for several values of the dimensionless gap parameter $\delta$
(that measures the
distance from the critical point): a) $\delta=0.5$, b) $\delta=0.1$,
c) $\delta=0.05$ and d) $\delta=0.0$.}
\label{fig2}
\end{figure}

\begin{figure}
\caption{Plot of the two parameter universal scaling function
(real part) of the regular contribution to the frequency dependent 
conductivity in the quantum critical regime.
\label{fig3}}
\end{figure}

\begin{figure}
\caption{Same as in Fig. 3 for the imaginary part}
\label{fig4}
\end{figure}
\end{document}